\documentclass[12pt,preprint]{aastex}

\def\gta{\ifmmode {\mathbin{\lower 3pt\hbox   
    {$\,\rlap{\raise 5pt\hbox{$\char'076$}}\mathchar"7218\,$}}}
    \else {${\mathbin{\lower 3pt\hbox
    {$\rlap{\raise 5pt\hbox{$\char'076$}}\mathchar"7218\,$}}}
    $}\fi}
\def\lta{\ifmmode {\,\mathbin{\lower 3pt\hbox   
    {$\,\rlap{\raise 5pt\hbox{$\char'074$}}\mathchar"7218\,$}}}
    \else {${\mathbin{\lower 3pt\hbox
    {$\rlap{\raise 5pt\hbox{$\char'074$}}\mathchar"7218\,$}}}
    $}\fi}

\begin{document}

\title{Probing General Relativity With Mergers of Supermassive and
Intermediate-Mass Black Holes}

\author{M. Coleman Miller}
\affil{Department of Astronomy, University of Maryland\\
       College Park, MD  20742-2421\\
       miller@astro.umd.edu}

\begin{abstract}

Recent observations and stellar dynamics simulations suggest that $\sim
10^3\,M_\odot$ black holes can form in compact massive young star
clusters.  Any such clusters in the bulge of their host galaxy
will spiral to the center within a few hundred million years, where
their intermediate-mass black holes are likely to merge eventually
with the galaxy's supermassive black hole.  If such mergers are
common, then future space-based gravitational wave detectors such as
the Laser Interferometer Space Antenna will detect them with such a
high signal to noise ratio that towards the end of the inspiral the
orbits will be visible in a simple power density spectrum,
without the need for matched filtering.  We discuss the astrophysics
of the inspiral of clusters in the nuclear region of a galaxy and the
subsequent merger of intermediate-mass with supermassive black holes.  We
also examine the prospects for understanding the spacetime geometry of
rotating black holes, based on phase connection of the strong signals
visible near the end of these extreme mass ratio inspirals.

\end{abstract}


\section{Introduction}

Observations of many star-forming galaxies show that a common mode of star
formation involves the production of young massive star clusters (or
``super star clusters"), which might have masses $\sim$few$\times
10^5\,M_\odot$ with half-mass radii of $\sim$few~pc (e.g., van den
Bergh 1971 and numerous subsequent papers; see Ma\'iz-Appell\'aniz 2001
for a review of the structural parameters of such clusters).  According to
recent N-body simulations (Ebisuzaki et al. 2001; Portegies Zwart \&
McMillan 2002; Portegies Zwart et al. 2004), Monte Carlo simulations
(G\"urkan, Freitag, \& Rasio 2004), and semi-analytic treatments (Mouri \&
Taniguchi 2002b), when such a cluster is compact enough it can evolve
dynamically in such as way as to produce runaway collisions in its center
in the few million years before the most massive stars explode. Such
collisions could produce a black hole of several hundred solar masses, and
subsequent dynamical processes could add additional mass to the black hole
(Miller \& Hamilton 2002a,b; Mouri \& Taniguchi 2002a; G\"ultekin, Miller,
\& Hamilton 2004). As suggested by Ebisuzaki et al. (2001), clusters of
this type that start close enough to the center of their host galaxy will
sink to the center within a few billion years, where they will eventually
release their black hole.  Rough estimates (see \S~3) suggest that a few
times per year, a merger of such an intermediate-mass black hole (IMBH)
with the central supermassive black hole (SMBH) will be detected by
space-based gravitational wave detectors such as the {\it Laser
Interferometer Space Antenna} ({\it LISA}).

If these mergers occur, they will be ideal sources with which to probe the
spacetime geometry around rotating black holes (see the discussion in
Cutler \& Thorne 2002).  The mass ratio (typically $10^{3-4}$) is large
enough that the IMBH acts almost as a test particle, but the signal
strength is much larger than it is for mergers of stellar-mass black holes
with SMBH (Hughes 2001; Cutler \& Thorne 2002; Glampedakis, Hughes, \&
Kennefick 2002).  As a result, although
there is much greater uncertainty about the event rate for IMBH-SMBH
mergers than for mergers of stellar-mass and supermassive black holes
(thus design considerations for {\it LISA} should focus on the latter),
if even a single IMBH-SMBH merger is detected, then high-precision
constraints on gravitational radiation and the Kerr spacetime will be
possible with greatly simplified data analysis.

Here we discuss the dynamics and implications of such IMBH-SMBH mergers.
In \S~2 we describe the astrophysical scenario of an influx of IMBHs
into the center of a galaxy.  In \S~3 we make estimates of the strength
of the signal, and discuss data analysis in the {\it LISA} context.
We summarize in \S~4.

\section{Astrophysical Scenario}

Throughout this paper, we consider interactions of a supermassive
black hole of mass $M$ with one or more intermediate-mass black holes
of mass $\mu\ll M$.  We scale these masses by 
$10^6\,M_\odot$ and $10^3\,M_\odot$, respectively.  

If a super star cluster of mass $M_{\rm cl}$ is embedded in a much
lower density stellar environment, it will act dynamically as a single
object.  Adapting equation (7-26) of Binney \& Tremaine (1987) and
equation (2) of Ebisuzaki et al. (2001), the
dynamical friction time for such a cluster to sink from a distance $r$
to the center of a galaxy with three-dimensional velocity dispersion 
$\sigma_{\rm gal}$ is
\begin{equation}
t_{\rm df}=(1.65/\ln\Lambda)(r^2\sigma_{\rm gal}/GM)\approx 
4\times 10^8~{\rm yr}(\sigma_{\rm gal}/100~{\rm km~s}^{-1})
(r/100~{\rm pc})^2(10^5\,M_\odot/M_{\rm cl})\; ,
\end{equation}
where $\ln\Lambda$ is the Coulomb logarithm.
Therefore, a cluster will be able to sink to the center within much
less than a Hubble time if it starts anywhere within the inner few
hundred parsecs of its host galaxy.

From this point, we expect the following sequence: (1)~the cluster
sinks until it is stripped or tidally disrupted, thus releasing its
IMBH, (2)~the IMBH sinks rapidly until the stellar mass interior to it
is less than the mass of the IMBH, (3)~the orbital radius of the IMBH
around the SMBH shrinks via interactions with stars, as long as the
relaxation time for the surrounding stars is less than a Hubble time,
and (4)~the IMBH either merges with the SMBH due to interactions with
stars followed by inspiral caused by gravitational radiation, or one or
more additional IMBHs settle to the center and interact dynamically,
causing mergers.  We now discuss each of these steps.

{\it Cluster mass loss.}---As the cluster sinks, it can lose stars 
in several ways (a similar discussion in the context of stars at
the Galactic center is in Hansen \& Milosavljevic 2003).  The first
is tidal stripping.  That is, if the cluster mass and radius are
$M_{\rm cl}$ and $R_{\rm cl}$, respectively, then the outer portions 
of the cluster will be stripped away when the cluster
is a distance $r<r_{\rm tide}$ from the center of the galaxy, where
the tidal radius $r_{\rm tide}$ is given by
\begin{equation}
r_{\rm tide}=\left[3M(<r_{\rm tide})/M_{\rm cl}\right]^{1/3}R_{\rm cl}\; .
\end{equation}
Observations of the central regions of many galaxies suggest that the
velocity dispersion is relatively constant (K. Gebhardt, personal
communication).  This is therefore consistent with an isothermal density
profile, in which $M(<r)=2\sigma^2 r/G$, where $\sigma$ is the
three-dimensional velocity dispersion (see equation 4-123 of Binney \&
Tremaine 1987). Rewriting, we find that the tidal
radius is
\begin{equation}
r_{\rm tide}=\left[6\sigma^2/(GM_{\rm cl}/R_{\rm cl})\right]^{1/2}R_{\rm cl}\; ,
\end{equation}
or about $10-20R_{\rm cl}$ for $M_{\rm cl}\sim{\rm few}
\times 10^5\,M_\odot$ and a half-mass radius $R_{\rm cl}\sim 
{\rm few}$~pc.  This will typically allow the cluster to sink in
to $\sim 30-50$~pc, which it does within $\sim 10^8$~yr if it started 
at $\sim 100$~pc.  If we assume that the cluster itself has mass
distributed roughly as an isothermal sphere, then $M_{\rm cl}\propto
R_{\rm cl}$ and therefore the relaxation time scales as
$t_{\rm rel}\propto r^2/M\propto R_{\rm cl}$ because the tidal
radius scales as $R_{\rm cl}$.

However, the cluster itself will also evolve dynamically.  From
the Pryor \& Meylan (1993) catalog of Galactic globular clusters,
the typical half-mass relaxation time for a globular is $\sim 10^{8-9}$~yr.
For a cluster with $N$ stars and a crossing time of $t_{\rm cross}
=R_{\rm cl}/\sigma_{\rm cl}$ (where $\sigma_{\rm cl}$ is the 
three-dimensional velocity
dispersion of the cluster), the cluster relaxation time
is $t_{\rm rel,cl}\approx (0.1N/\ln N)t_{\rm cross}$ (e.g., Binney \&
Tremaine 1987).  For an isothermal
sphere, $N\propto R_{\rm cl}$ and $t_{\rm cross}\propto R_{\rm cl}$.
Thus, $t_{\rm rel,cl}\propto R_{\rm cl}^2$.  

Once tidal stripping of the cluster begins, therefore, the cluster
relaxation time will decrease faster than the dynamical friction time.
When $t_{\rm rel,cl}<t_{\rm df}$, the cluster will disperse and the
IMBH will be on its own.  For typical masses and radii of clusters, the
above simplified treatment would suggest that this will happen when
$r\sim 10$~pc.  Given that clusters that form IMBHs tend to have
short relaxation times, there could be a concern that these clusters
would disrupt earlier.  However, simulations by Kim, Figer, \& Morris (2004)
and by A. G\"urkan \& F. Rasio (in preparation) support the suggestion
of Hansen \& Milosavljevic (2003) that the presence of an IMBH in the
center of a cluster increases the velocity dispersion of the stars and
hence their relaxation time.  Therefore, it is found numerically
that in fact the IMBH is released at $\sim$few~pc. Thus equation (1)
suggests that the IMBH will take $\lta 10^8~{\rm yr} (\sigma/100~{\rm
km~s}^{-1})(10^3\,M_\odot/\mu)$ to sink to the center. 
Clusters that start within $\sim 100$~pc of the center will be able to
deliver their central intermediate-mass black holes to the center
within a few hundred million years.

For completeness, we now discuss another way in which a cluster
could theoretically be dispersed.  Given that the velocity dispersion
of stars in a galactic bulge is much greater than the velocity dispersion
of stars in a cluster, passage of bulge stars through the cluster will
soften the cluster somewhat, and will eventually cause it to evaporate.
One can show, however, that this effect is unimportant.  From 
Binney \& Tremaine (1987, equation 4-6a), the typical change in squared
transverse velocity of a particle of mass $m$ going at speed $v$
through a cluster of $N$ particles of mass $m$ within a radius $R$ is
\begin{equation}
\Delta v_\perp^2\approx 8N(Gm/Rv)^2\ln\Lambda
\end{equation}
where $\ln\Lambda\sim 10-20$ is a Coulomb logarithm.
If we use $v=\sigma_{\rm gal}$ and assume a cluster velocity dispersion
of $\sigma_{\rm cl}^2\approx GNm/R$, then this becomes
\begin{equation}
\Delta v_\perp^2\approx (8\ln\Lambda/N)\sigma_{\rm cl}^2(\sigma_{\rm cl}/
\sigma_{\rm gal})^2\; .
\end{equation}
Because these are softening interactions, we will assume that the
energy of the cluster is always {\it increased} by ${1\over 2}m
\Delta v_\perp^2$.

As these are fast interactions, there is little gravitational
focusing and hence the mass per time interacting with the cluster
is simply $\rho(\pi R_{\rm cl}^2)\sigma_{\rm gal}$.  For an isothermal sphere,
$\rho=\sigma_{\rm gal}^2/(2\pi r^2 G)$ at distance $r$ from the center.
The change in energy per time is then
\begin{equation}
\begin{array}{rl}
dE/dt&={1\over 2}{\sigma_{\rm gal}^2\over{2\pi r^2 G}}
(\pi R_{\rm cl}^2)\sigma_{\rm gal}{8\ln\Lambda\over{N}}\sigma_{\rm cl}^2
\left(\sigma_{\rm cl}/\sigma_{\rm gal}\right)^2\\
&={2\sigma_{\rm gal}\over{r^2G}}R_{\rm cl}^2{\ln\Lambda\over{N}}
\sigma_{\rm cl}^4\; .\\
\end{array}
\end{equation}
The total binding energy of a singular isothermal sphere is
\begin{equation}
E=\int_0^{R_{\rm cl}}{GM(<r)\over{r}}\rho dV=2\sigma_{\rm cl}^4R_{\rm cl}/G\; .
\end{equation}
Therefore, the softening time is
\begin{equation}
t_{\rm soft}=E/(dE/dt)={r^2N\over{\sigma_{\rm gal}R_{\rm cl}\ln\Lambda}}\; .
\end{equation}
The Coulomb logarithms for $t_{\rm soft}$
and for $t_{\rm df}$ will be different in general, but probably not by more
than a factor of a few ($\ln\Lambda$ for softening is likely to be of
order 10-15, but for dynamical friction is probably 3-5; see Spinnato
et al. 2003).  Therefore, as an approximation we can effectively cancel 
the Coulomb logarithms when we take the ratio:
\begin{equation}
t_{\rm soft}/t_{\rm df}\sim N(GM_{\rm cl}/R_{\rm cl})/\sigma_{\rm gal}^2
\sim N(\sigma_{\rm cl}/\sigma_{\rm gal})^2\gg 1\; .
\end{equation}
For example, if $N=10^6$, $\sigma_{\rm gal}=100$~km~s$^{-1}$, and
$\sigma_{\rm cl}=10$~km~s$^{-1}$, then $t_{\rm soft}\sim 10^4
t_{\rm df}$.  Softening by interactions with bulge stars can always be
neglected in comparison with other effects.

{\it Initial inspiral of the IMBH.}---After the cluster disrupts, 
the IMBH itself will spiral in independently.
As a first stage, it will spiral in to where the mass
interior to it is not much less than the mass of the IMBH itself.  For a
stellar number density of $10^6$~pc$^{-3}$, this implies a distance of
$\sim 0.05$~pc, but for a higher density it will be less.  For example,
Hansen \& Milosavljevic quote the Genzel et al. (2003) density profile
of the central cusp of the Galaxy as implying $M(<r)=1.3\times 10^4\,M_\odot
(r/0.04~{\rm pc})^{1.63}$, where 1"=0.04~pc at 8~kpc.  This implies
a higher density, so that the rapid inspiral of a $10^3\,M_\odot$ IMBH
will occur down to a separation of $\sim 0.01$~pc.  From above, the
inspiral of the IMBH will start from a few parsecs, hence it will come in on 
a timescale
of $\lta 10^8$~yr for typical densities and velocity dispersions.

{\it Long-term inspiral of the IMBH.}---Further settling of the IMBH
requires that it interact with a significant mass in stars.  If the stars
have fully isotropized orbits, this is easy: for a strongly
gravitationally focused encounter with a binary of total mass $M$ and
semimajor axis $a$ the cross section is $\Sigma=\pi a (2GM/\sigma^2)$ and
the timescale of interaction is $\tau=1/(n\Sigma\sigma)$, which is much
less than a year for typical masses, velocities, and densities.

However, stars that interact with the IMBH-SMBH binary are eventually
thrown out of the system, so the bottleneck is the time needed for other
stars to diffuse into the required orbital phase space.  This ``loss cone"
of stars could cause supermassive black hole binaries to stall in their
inspiral, before they get close enough for gravitational radiation to be
significant (Begelman, Blandford, \& Rees 1980; see Sigurdsson \& Rees 1997,
Milosavljevic \& Merritt 2003, Sigurdsson 2003, and 
Makino \& Funato 2004 for recent discussions).  For IMBHs, this is not
likely to be a problem.  As discussed by Yu \& Tremaine (2003), once
the original contingent of stars is ejected from the loss cone, the
system will settle into a state in which the rate of diffusion of 
stars into the loss cone is balanced by the rate at which they are
ejected by interaction with the IMBH-SMBH binary.  From equation (38)
of Yu \& Tremaine (2003), the hardening timescale for a black hole
binary of total mass $M$ is
\begin{equation}
t_h\approx 6\times 10^9~{\rm yr}(M/3.5\times 10^6\,M_\odot)(1~M_\odot/m_*)
(2\times 10^{-4}~{\rm yr}^{-1}/n_{\rm diff})\; ,
\end{equation}
where $m_*$ is the typical stellar mass in the central regions and
$2\times 10^{-4}~{\rm yr}^{-1}$ is a characteristic value for the
diffusion rate $n_{\rm diff}$ into the loss cone.  Therefore, depending
on the details of the stellar distribution, the orbital radius of the
IMBH could be reduced by several e-foldings in a Hubble time, especially
if the SMBH has $M\lta 10^6\,M_\odot$.
This process could be enhanced slightly because stars that interact
with the IMBH will typically not be ejected entirely from the core, hence
they will return for several interactions (Milosavljevic \& Merritt 2003).
In addition, gas dynamical friction from molecular clouds (see, e.g.,
Ostriker 1999) can shrink the orbit further.

{\it Final merger with the SMBH.}---By the time $a<10^{-3}$~pc,
gravitational radiation can be important for an IMBH-SMBH binary.  The
timescale to merger is
\begin{equation}
\tau_{\rm GR}\approx 10^{12}~{\rm yr}(\mu/10^3\,M_\odot)^{-1}
(M/10^6\,M_\odot)^{-2}(a/0.001~{\rm pc})^4(1-e^2)^{7/2}
\end{equation}
(from, e.g., Peters 1964) where $e$ is the orbital eccentricity.  Thus,
if $a<0.0003$~pc or the eccentricity is high, merger can happen within
a Hubble time.

Therefore, in contrast what might be the case for two supermassive black
holes in a binary (Begelman et al. 1980), it is unlikely that there is a
hang-up  problem for an IMBH-SMBH binary.  The difference is that an
IMBH-SMBH binary at a given separation has a much smaller binding
energy than a binary with two supermassive black holes. Hence, the 
stars that are ejected or displaced in the process of
hardening the binary come from a relatively smaller volume, in which the
relaxation time is short enough to repopulate the loss cone. Given that
each IMBH by assumption brings with it several hundred thousand new
stars, there will always be a fresh set of stars to supply dynamical
friction.  It is therefore possible that tens or even hundreds of IMBHs
could be brought in sequentially, each merging with the SMBH before the
next IMBH arrives.

Note that this situation is dramatically different from the processes
for mergers of stellar-mass black holes with supermassive black holes.
In that case, the dynamical friction time for stellar-mass black holes
is much too long to get to the center in a Hubble time.  As a result,
only rare scatters of stellar-mass black holes into extremely high
eccentricity orbits, followed by capture onto the SMBH by release of
energy in gravitational radiation, can lead to a merger (e.g., Freitag
2003; Sigurdsson 2003).  In contrast,
the scenario we describe for IMBHs leads to sinking of the IMBH towards
the center on a relatively short timescale.  Gravitational radiation
capture of black holes on hyperbolic orbits is not necessary.

If the timescale for dynamical friction and merger is longer
than the timescale for the next IMBH to sink in (e.g., because the
stellar number density at the center is much less than we have
assumed), then a few IMBHs will interact with each other as they orbit
the SMBH.  This will lead to instabilities in the orbits.  The exact
criterion for instability depends on mass ratios and eccentricities
(e.g., see Mardling \& Aarseth 2001 for a comparable-mass binary
orbited by a tertiary of arbitrary mass), but if orbits of particles
approach each other within a few tens of percent of their orbital radii
then instability usually results.  

Once this occurs, the orbiting IMBHs will interact with each other until
either (1)~secular resonances drive the inner IMBH close enough to the
SMBH that the pair merges because of gravitational radiation (a situation
that preliminary simulations suggest may be surprisingly common), or
(2)~one or several IMBHs are ejected, implying by energy conservation
that the inner one or several IMBHs are driven closer to the SMBH.  In
the latter case, simulations must be performed to determine the
efficiency of this process, that is, the average number of IMBHs ejected
for each one that merges.  If simulations of stellar-mass black holes
around an IMBH are a guide, then ejections may be dominant (see, e.g.,
Baumgardt, Makino, \& Ebisuzaki 2004).  However, the dynamics of 
SMBH-IMBH systems could be different in several important ways.
For example, if an IMBH is ejected from the
core but not the entire bulge, its periapse is still of order the
IMBH-SMBH binary semimajor axis, so barring significant deflection during
its orbit it could interact again on the next pass.  In addition,
although the IMBH-SMBH mass ratio is small enough to prevent ejection of
the binary, if the inner region has been evacuated of stars because of
prior interactions then the small binary kick due to IMBH ejection will
cause the binary to move significantly, to where it can interact with
more stars and harden further. Numerical details of the interactions also
need to be computed to estimate quantities such as the eccentricity in
the sensitivity band of a particular gravitational radiation detector,
and to determine whether two IMBHs might pass close enough to each other
to form bound pairs by the loss of energy to gravitational radiation,
leading to IMBH-IMBH mergers (D. Hamilton, personal communication).

For the purposes of this paper, however, the main point is that the IMBHs
are expected to merge with the central SMBH eventually, rather than
stalling or being ejected.  As we now discuss, this is a high mass ratio
merger (and hence comparatively easy to calculate) with a large enough
signal to noise ratio that it will be possible to detect it near the end
of inspiral in just a few cycles, requiring very few templates.

\section{Detection of IMBH-SMBH Gravitational Radiation}

The information content of the signal from an IMBH-SMBH binary depends
on the signal to noise ratio.  To compute the signal strength, we assume
for simplicity that the binary is nearly circular by the time it enters
the sensitivity band of an instrument such as {\it LISA}; we will discuss
the possibility of an eccentric binary in \S~4.

The rest-frame frequency of gravitational radiation from a 
nearly circular binary a time $T_{\rm merge}$ from merger is
(see Peters \& Mathews 1963; Peters 1964 for the basic equations)
\begin{equation}
f_{\rm GW,rest}=7\times 10^{-4}~{\rm Hz}(\mu/10^3~M_\odot)^{-3/8}
(M/10^6~M_\odot)^{-1/4}(T_{\rm merge}/1~{\rm yr})^{-3/8}\; ,
\end{equation}
at which point the orbital semimajor axis in units of the gravitational
radius $r_g=GM/c^2$ is
\begin{equation}
a/r_g=19(\mu/10^3\,M_\odot)^{1/4}(M/10^6\,M_\odot)^{-1/2}
(T_{\rm merge}/1~{\rm yr})^{1/4}\; .
\end{equation}
If the source is at a redshift $z$ then the observed frequency is
$f_{\rm obs}=f_{\rm GW,rest}/(1+z)$.
From, e.g., Schutz (1997), the dimensionless amplitude of a circular 
binary at a line of sight comoving distance $D_M$, averaged over all 
observer angles, is
\begin{equation}
\begin{array}{rl}
h&=2^{2/3}(4\pi)^{1/3}G^{5/3}c^{-4}f_{\rm GW,rest}^{2/3} \mu M^{2/3}/D_M\\
&=1.3\times 10^{-21}(\mu/10^3~M_\odot)^{3/4}(M/10^6~M_\odot)^{1/2}
(T_{\rm merge}/1~{\rm yr})^{-1/4}(3~{\rm Gpc}/D_M)\; .\\
\end{array}
\end{equation}
At the innermost stable circular orbit (ISCO) for a nonrotating SMBH,
$a_{\rm ISCO}=6GM/c^2$, the amplitude and rest-frame frequency are
\begin{equation}
\begin{array}{rl}
h_{\rm ISCO}&=1.4\times 10^{-20}(\mu/10^3\,M_\odot)(3~{\rm Gpc}/D_M)\\
f_{\rm ISCO}&=4.4\times 10^{-3}~{\rm Hz}(M/10^6\,M_\odot)^{-1}\; .\\
\end{array}
\end{equation}
Note that the amplitude at the ISCO is independent of $M$, because
$h\propto f^{2/3}M^{2/3}$ and $f\propto M^{-1}$.

The effective {\it LISA} noise includes contributions from the instrument and 
from unresolved binaries (e.g., see Larson, Hiscock, \& Hellings 2000
and 
http://www.srl.caltech.edu/$\sim$shane/sensitivity/MakeCurve.html).  
From $\sim 2\times 10^{-4}-2\times 10^{-3}$~Hz,
unresolved Galactic double white dwarf binaries exceed the instrumental
noise (e.g., Farmer \& Phinney 2003); from $\sim 2\times 10^{-3}-10^{-2}$~Hz,
in contrast, there will typically be one or zero double white dwarf binaries
in a $10^{-8}$~Hz bin, hence after several years of operation, it will
be possible to model individual binaries and subtract them from the
data stream.  Unresolved extragalactic double white dwarf binaries will,
however, continue to make a contribution.  The minimum total noise is
in the few mHz range, where the total one-sided spectral noise density at a 
signal to noise $S/N=10$ is
\begin{equation}
S_n(10\sigma)\approx 1.5\times 10^{-19}~{\rm Hz}^{-1/2},\qquad 
3\times 10^{-3}~{\rm Hz}<f_{\rm obs}<10^{-2}~{\rm Hz}\; .
\end{equation}

The time necessary to detect an SMBH-IMBH binary at S/N=10 is 
$T_{\rm obs}=\left[S_n(10\sigma)/h\right]^2$.
If $3\times 10^{-3}~{\rm Hz}<f_{\rm obs}<10^{-2}~{\rm Hz}$, then 
\begin{equation}
T_{\rm obs}\approx 1.2\times 10^4~{\rm s}(\mu/10^3\,M_\odot)^{-3/2}
(M/10^6\,M_\odot)^{-1}(T_{\rm merge}/1~{\rm yr})^{1/2}(D_M/3~{\rm Gpc})^2\; .
\end{equation}
Multiplying $f_{\rm obs}$ by $T_{\rm obs}$ gives the number of cycles 
in the time $T_{\rm obs}$:
\begin{equation}
N=8(1+z)^{-1}(\mu/10^3\,M_\odot)^{-15/8}(M/10^6\,M_\odot)^{-5/4}
(T_{\rm merge}/1~{\rm yr})^{1/8}(D_M/3~{\rm Gpc})^2\; .
\end{equation}
The minimum observational time and number of cycles are obtained when the
source is near the ISCO, which occurs in the most favorable frequency
band $3\times 10^{-3}~{\rm Hz}<f_{\rm obs}<10^{-2}~{\rm Hz}$ when the
redshifted mass $M(1+z)$ is between
$1.5\times 10^6\,M_\odot$ and $4.4\times 10^5\,M_\odot$.  At this point,
\begin{equation}
\begin{array}{rl}
T_{\rm obs,min}&=1200~{\rm s}(\mu/10^3\,M_\odot)^{-2}(D_M/3~{\rm Gpc})^2\\
N_{\rm min}&=5(1+z)^{-1}(\mu/10^3\,M_\odot)^{-2}(M/10^6\,M_\odot)^{-1}
(D_M/3~{\rm Gpc})^2\; .\\
\end{array}
\end{equation}
More generally, as in Figure~1, one can compute the minimum observation
time and number of cycles for S/N=10, $\mu=10^3\,M_\odot$, and any $M$, based
on the frequency at the ISCO and the projected total noise curve.
A prograde encounter with a rapidly rotating SMBH will go to higher
frequencies during its inspiral than will an encounter with a 
nonrotating SMBH.  This increases the energy released
in gravitational radiation and, importantly, increases the mass threshold
at which the observed signal is in the most sensitive frequency range
of the {\it LISA} band.  The numbers in Figure~1 are therefore conservative.

The expected rate of such events depends on a number of uncertain
astrophysical parameters.  In particular, it is clear that the low mass
end of SMBH (say $\lta 10^6\,M_\odot$) is of great importance.  Yu \& Lu
(2004) use the velocity dispersion data of Sheth et al. (2003) to
estimate that the comoving number density of black holes in this mass
range is $\sim$few$\times 10^{-3}$~Mpc$^{-3}$.  Out to $\sim 3$~Gpc
(where $z\approx 0.8$ so redshift corrections are moderate), the volume of
the universe is $\approx 10^{11}$~Mpc$^3$, implying $\sim {\rm
few}\times 10^8$ black holes in the required mass range.  If on average
$N_{\rm merge}$  IMBH-SMBH mergers per galaxy happen in $\sim
10^{10}$~yr, this implies  an overall rate of a few percent of $N_{\rm
merge}$ per year. 

The value of $N_{\rm merge}$ is highly uncertain.  The $M-\sigma$
relation (e.g.,Ferrarese \& Merritt 2000; Gebhardt et al. 2000; Merritt
\& Ferrarese 2001a,b; Tremaine et al. 2002) implies that the SMBH
typically contains $\sim 10^{-3}$ of the mass of the central bulge,
which means that $M_{\rm bulge}\sim 10^9\,M_\odot$ for $M\sim 10^6
\,M_\odot$.  If $\sim 10$\% of this mass was originally in the form of
young massive clusters (which later merged with the bulge), and if a few
tens of percent of such clusters form IMBHs, this suggests  $N_{\rm
merge}\approx 100$ over the lifetime of the galaxy.   This is consistent
with observations of  actively interacting galaxies such as M82, which
have hundreds of super star clusters younger than $10^8$~yr and
presumably have had many times that number over their lifetimes.  Note
that the total mass added by such mergers is much less than the mass of
an SMBH, hence this number of mergers is not in conflict with limits
based on the integrated light from quasars (Yu \& Tremaine 2002).  It is
therefore reasonable that there will be several IMBH-SMBH mergers
detectable with {\it LISA} during its few year lifetime.

The observable number and precision of inferences could change depending 
on the astrophysics involved.  For
example, if most massive clusters are formed at $z\sim 2$ in accordance
with the peak in the star formation history of the universe (e.g.,
Madau, Pozzetti, \& Dickinson 1997) and their IMBHs merge in
$<$1~Gyr with the SMBH, then most mergers are at a high enough redshift
that the frequencies are low and hence the S/N values are decreased.  
Even in this case, the signal strength could be large enough that elaborate
templates are unnecessary for detection (see \S~4).  If in
contrast the process of spiraling in and merging typically takes a few
billion years, mergers will be distributed over time and a
significant number of them will take place at low redshift when the S/N
is high in just a few cycles.  

The maximum distance at which an
IMBH-SMBH binary could be detected at $S/N>10$ (with perfect signal
processing) can be estimated from equation (15).  The line of sight
comoving distance saturates at high redshift (see, e.g., Peebles 1993,
chapter 13), to $\sim 10$~Gpc for cosmological parameters
$\Omega_M=0.27$, $\Omega_\Lambda=0.73$, and $H_0=72$~km~s$^{-1}$~Mpc$^{-1}$
(e.g., Spergel et al. 2003).  From equation (15), the amplitude and
observed frequency near the ISCO are then $h_{\rm ISCO}\approx 4\times 10^{-21}
(\mu/10^3\,M_\odot)$ and $f_{\rm ISCO,obs}=4.4\times 10^{-3}~{\rm Hz}
(M/10^6\,M_\odot)$, where $\mu$ and $M$ are measured in the rest frame.
From Larson et al. (2000), $S/N=10$ in a one year {\it LISA} integration
crosses an amplitude of $4\times 10^{-21}$ at a frequency of
$\approx 2\times 10^{-4}$~Hz, including white dwarf noise, hence
a $1000\,M_\odot-10^6\,M_\odot$ binary could be observed out to a
redshift $z\approx 20$ at $S/N=10$ in a one year integration.

\section{Discussion}

The scenario discussed in this paper relies on still uncertain details
of the production and distribution of intermediate-mass black holes
(see Miller \& Colbert 2004 for a discussion of formation mechanisms,
and of issues such as wind losses in the formation of high mass stars).
Here we have focused on the particular idea that IMBHs are formed in
runaway collisions in clusters.  Other formation mechanisms have
different implications.  For example, Madau \& Rees (2001) propose that
IMBHs form from the evolution of solitary nearly zero metallicity
(Population III) stars in the early universe.  In such a case, hierarchical
merging of minihalos could produce multiple IMBH-SMBH mergers in the
high redshift universe.  However, at this point too little is known about
such scenarios to make informed estimates of rates.  Our main point is
that if even a few IMBH-SMBH mergers are detected they will be useful as
uniquely precise tests of strong gravity.

To see this, consider first the inspirals of stellar-mass black holes into
supermassive black holes.  These are promising as probes of the Kerr
spacetime, but a difficulty is that the waves are expected to be weak
enough that thousands of orbits are required to achieve a reasonable signal
to noise (e.g., Barack \& Cutler 2004).  As a result, a very large number
of templates are required to detect the signal, which could make analysis
difficult.  In contrast, if IMBH-SMBH mergers occur a few times per year,
their signal strengths will lead to detections within just a few orbits,
near the end of inspiral.  As a result, as we now show, only standard
Fourier transforms are needed rather than any elaborate templates.

Consider the time for a nearly circular
orbit to merge (Peters \& Mathews 1963; Peters 1964):
\begin{equation}
T_{\rm merge}\approx 6\times 10^{17}~{\rm yr}(M_\odot^3/\mu M^2)
(a/1~{\rm AU})^4\; .
\end{equation}
For observation times $T_{\rm obs}\ll T_{\rm merge}$, the change in
gravitational wave frequency is $\Delta f\sim (T_{\rm obs}/T_{\rm merge})
f_{\rm obs}$.  The frequency resolution is $\delta f=1/T_{\rm obs}$, so if
$\delta f>\Delta f$ the signal shows up as a single peak in a power density
spectrum.  Therefore, if one observes for a coherence time $T_{\rm
coh}=\left(T_{\rm merge}/f_{\rm obs}\right)^{1/2}$ (such that $\delta
f=\Delta f$), one has the maximum possible power in a single peak in a
power density spectrum.  In Figure~2 we show the signal to noise ratio for
circular orbits over a coherence time for different unredshifted SMBH
masses (assuming in each case $\mu=10^3\,M_\odot$), for observed
frequencies from $10^{-4}$~Hz to $f_{\rm ISCO}/(1+z)$, where we assume
$D_M=3$~Gpc and therefore $z=0.8$.  From this figure we see that if
$M<10^6\,M_\odot$ then a circular signal will be detectable with S/N$>$10
in a coherence time near the end of inspiral.  If there are closer mergers,
say with $D_M=1$~Gpc, the signal to noise could be as large as hundreds.

As a result, if IMBH-SMBH mergers occur, then during the end of inspiral
they are detectable without modeling.  At earlier times this is not the
case, but it will be possible to use the late-time detections to work
backwards and determine the full set of orbital parameters 
by connecting the phases of the individual segments.  It will also be
possible to establish very precise initial conditions for numerical
modeling of the merger phase.  Some idea of the precision with which
parameters will be estimated for such a merger (after fitting a
year-long wave train) can be obtained from Tables II and III of
Barack \& Cutler (2004).  Linear scaling from these results is
not appropriate, given correlations between parameters, but the
much greater signal to noise ratio of IMBH-SMBH mergers (thousands
instead of tens) suggests that, for example, the redshifted masses
and the dimensionless angular momentum of the SMBH will be estimated
to fractional precisions of better than $10^{-5}$.

If the orbit is eccentric, or if other effects (e.g., pericenter
precession or Lense-Thirring precession) produce peaks separated in
frequency by more than $1/T_{\rm coh}$ from the main peak, then the
analysis is complicated somewhat.  However, these frequencies will also
remain stable over $T_{\rm coh}$, so with high signal to noise one will
be able to detect each of these peaks independently and model the
changes in eccentricity, orbital inclination, and so on by building
up the full wave train.

As with mergers of stellar-mass with supermassive black holes, the
orbits of IMBHs into SMBHs will map out the Kerr spacetime and test
the no-hair theorem (e.g., Ryan 1997).  In addition, we point out that
the rate of inspiral (and decay of eccentricity if this is
nonnegligible) will provide a strict testbed for theoretical
predictions of the flux and angular momentum functions in strong
gravity.  For example, for a $10^3\,M_\odot-10^6\,M_\odot$ binary, the
total  S/N is $>10^4$ for the portion of the orbit inside of $10\,M$, so 
high-order contributions can be inferred empirically.

In future work we will proceed in two directions.  First, we will
explore the quantitative constraints on current post-Newtonian models
that are possible from detection of an IMBH-SMBH merger.  Second, we
will investigate astrophysical scenarios in which the orbit would
have significant eccentricity when the source is in the detectability
band of {\it LISA}.  Such eccentric orbits could arise from the
scenario we discuss here, or from plunge orbits as in
stellar-mass/supermassive mergers, or possibly from other
mechanisms.  If such scenarios are plausible, there is substantial
extra information to be gleaned. Virtually all templates constructed
so far are specialized for ground-based detections of high-frequency
waves, and hence assume that the orbits would have nearly
circularized by the time the gravitational waves entered instrumental
bands (see, e.g., Damour, Iyer, \& Sathyaprakash 2002 for an update
to 3.5PN order).  The lack of analysis of post-Newtonian expansions
of eccentric orbits means that observed eccentricity decay will at
least provide self-consistency checks, and possibly constrain
additional PN parameters beyond those that have been investigated
currently. Even if the orbits turn out to be mostly circular, there
is a wealth of data that could be extracted from mergers of
supermassive and intermediate-mass black holes.

\acknowledgements

We appreciate the hospitality of the Center for Gravitational Wave Physics
at Penn State during the workshop in which some of these ideas were
explored.  We are grateful for helpful discussions with Alessandra
Buonanno, Marc Freitag, Kayhan G\"ultekin, Brad Hansen, Fred Rasio, and
Steinn Sigurdsson.   The paper also benefited from an unusually helpful
referee's report. This work was supported in part by NASA grant NAG
5-13229.

\newpage

\centerline{\plotone{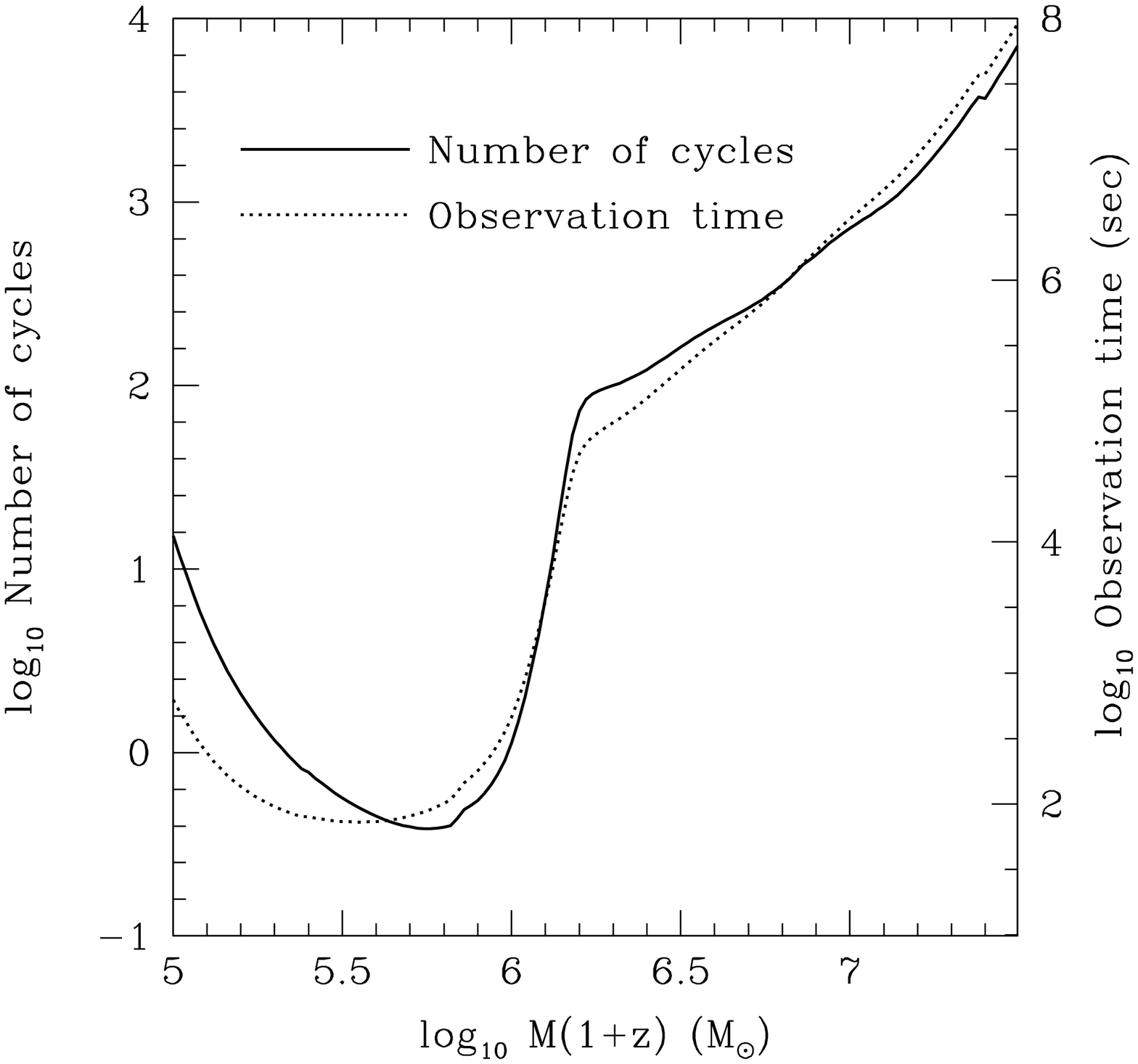}}
\figcaption[]{Minimum observation time (dotted line) and corresponding number
of gravitational wave cycles (solid line) required to get S/N=10
at the innermost stable circular orbit from a circular binary
of total mass $M$ and reduced mass $\mu=10^3\,M_\odot$, at a line of sight
comoving distance of $D_M=3$~Gpc.  This figure indicates the time and
cycles needed if the IMBH were to be fixed in an orbit at the ISCO; in
reality, the IMBH will typically spend several months at frequencies
comparable to $f_{\rm ISCO}$.  Therefore, if
$M(1+z)\lta{\rm few}\times 10^6\,M_\odot$, it is possible to achive
a high signal to noise in a very short time with an IMBH-SMBH binary.}

\newpage

\centerline{\plotone{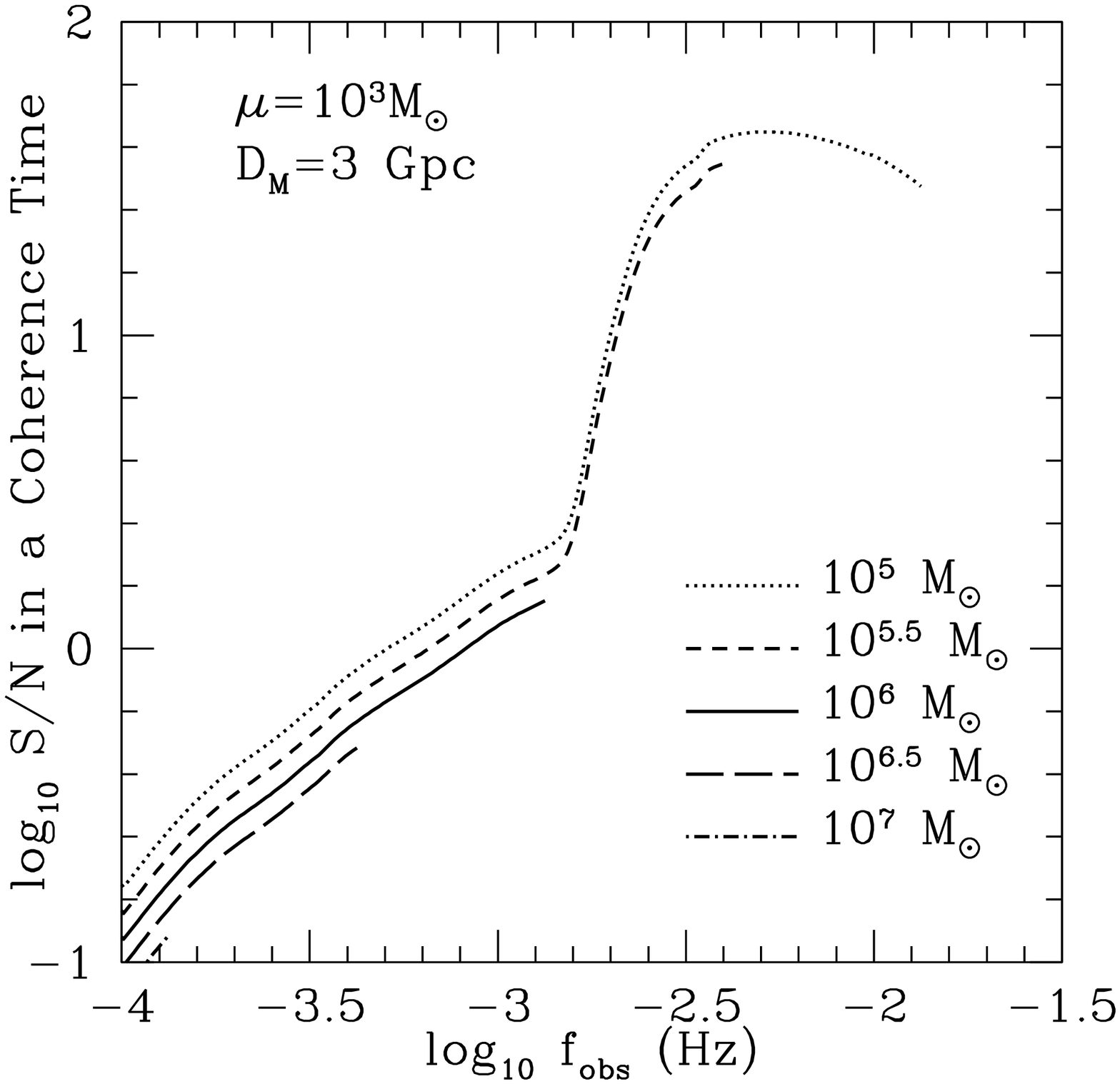}}
\figcaption[]{Signal
to noise in a coherence time (see text) for a binary at a line of sight
comoving distance $D_M=3$~Gpc that has an IMBH mass $\mu=10^3\,M_\odot$
and several possible total masses.  Here both instrumental noise and white
dwarf noise are included.  We plot S/N versus frequency, from
$f_{\rm obs}=10^{-4}$~Hz to the observed frequency at the innermost stable
circular orbit (we assume a redshift $z=0.8$ at 3~Gpc).  This is the
maximum signal obtainable in a simple power density spectrum.  For
$M<10^6\,M_\odot$, the signal will be detected strongly in a coherence
time, greatly simplifying data analysis.}

\end{document}